# A Simple and Efficient Non-DFT-Based Machine Learning Interatomic Potential to Simulate Titanium MXenes

Luis F. V. Thomazini[1,2] and Alexandre F. Fonseca[2]*

[1] *Instituto Federal de Educação, Ciência e Tecnologia de São Paulo, Campus Caraguatatuba, Caraguatatuba, SP CEP 11665-071, Brazil.*
[2] *Universidade Estadual de Campinas (UNICAMP), Instituto de Física Gleb Wataghin, Departamento de Física Aplicada, Campinas, São Paulo, 13083-859, Brazil.*

*Corresponding author: afonseca@ifi.unicamp.br

## ABSTRACT

*Titanium MXenes are two-dimensional inorganic structures composed of titanium and carbon or nitrogen elements, with distinctive electronic, thermal and mechanical properties. Despite the extensive experimental investigation, there is a paucity of computational studies at the level of classical molecular dynamics (MD). As demonstrated in a preceding study, known MD potentials are not capable of fully reproducing the structure and elastic properties of every titanium MXene. In this study, we present a simply trained, but yet efficient, non-density functional theory-based machine learning interatomic potential (MLIP) capable of simulating the structure and elastic properties of titanium MXenes and bulk titanium carbide and nitride with precision comparable to DFT calculations. The training process for the MLIP is delineated herein, in conjunction with a series of dynamical tests. Limitations of the MLIP and steps towards improving its efficacy to simulate titanium MXenes are discussed.*

## INTRODUCTION

MXenes are a class of two-dimensional (2D) materials composed by the formula $M_{n+1}X_n$, where M represents an early transition metal and X can be carbon or nitrogen [1]. Specifically, titanium carbides ($Ti_{n+1}C_n$) were the first MXenes synthesized [2], followed by titanium nitrides ($Ti_{n+1}N_n$) a few years later [3]. The properties of these materials [4,5] have encouraged research in a variety of applications [4-7].

A substantial body of research has been dedicated to the computational analysis of MXenes [8]. In a recent study [9], a comparative analysis was conducted of the accuracy in the determination of the structural and elastic properties of $Ti_{n+1}C_n$ and $Ti_{n+1}N_n$ ($n$ = 1, 2 and 3) MXenes between three well-known classical molecular dynamics (MD) potentials and density functional theory (DFT) calculations. The MD potentials examined in the study include the Charge-Optimized Many-Body (COMB3), the Reactive Force Field (REAXFF) and the Modified Embedded Atom Method (MEAM). The study found two main points. Only one ReaxFF force field [10] ([11]) was shown to reasonably simulate the titanium carbide (nitride) MXenes. A MEAM force field was shown to be suitable for simulating $Ti_3C_2$, $Ti_4C_3$, $Ti_3N_2$ and $Ti_4N_3$ [12]. The remaining MD potentials and force fields significantly under or overestimate the structural and/or elastic properties of titanium MXenes [9].

In this study, the methods in machine-learned interatomic potentials (MLIP) for classical MD simulations [13] are used to develop and evaluate two MLIP models, one for simulations of titanium carbide and the other for titanium nitride MXenes. Notably, the structures employed in the MLIP training methodologies were not obtained from DFT calculations, but rather from optimized structures using the aforementioned REAXFF force fields [10,11]. Despite the non-use of DFT structures, the two novel MLIPs are shown to reproduce the structural and elastic properties of all titanium MXenes with precision comparable to that of DFT calculations, as well as to provide satisfactory dynamic results. A subsequent discussion will address the surprising results as well as the limitations of these simple MLIPs for simulating titanium MXenes.

In the next sections, the theoretical framework and computational methodologies employed in the tests will be delineated, followed by the presentation and discussion of the results. In the final section, a summary of the primary conclusions is provided.

## THEORY AND SIMULATION DETAILS

The MLIP under development is based on the Moment Tensor Potential (MTP) [14], a class of mathematical descriptors that capture the local environment of atoms. Unlike classical interatomic potentials, MTP's energy can be adjusted from a reference database, thereby reducing computational cost and providing results comparable to that of DFT. This renders MTP an efficient alternative for large-scale more precise simulations [15]. MTP's description and implementation in the software package LAMMPS [16] are detailed in Ref. [17]. Here, we present the MLIP parameters, structures, and tests.

The present MLIP training process consists of three steps: 1. slowest energy structure; 2. high temperature endurance; and 3. elastic properties. The training steps provide the structural configuration (i.e., the atom positions) and the total energy of the configuration. The stress tensor and forces acting on each atom are not provided or made equal to zero for optimized structures. The level of MTP selected was 6, which is considered one of the simplest levels corresponding to vectorial descriptors. The radial basis functions are of the Chebyshev type, the size of the radial basis is 8, and the minimum and maximum distances are set to 1 and 4 Å, respectively. For all training steps, 2 × 2 supercells of each structure (and 2 × 2 × 2 for the bulk ones) were selected. The remaining parameters can be directly observed in the MLIP files which are available in the supplementary information.

1. Slowest energy structure. In this training step, the structures of all titanium carbide (nitride) MXenes and that of the bulk TiC (TiN) are used. The data for these structures include atom positions, sizes for periodic boundary conditions, null stress tensor, null force acting on each atom, and total energy. Atom positions were obtained from energy minimizations performed in LAMMPS with the best REAXFF force fields [10,11], according to our previous study [9]. The cohesive energies of the structures were chosen to be 6.08, -6.52, -6.71 and -7.19 eV/atom for $Ti_2C$, $Ti_3C_2$, $Ti_4C_3$ and TiC, respectively (-6.00, -6.14, -6.64 and -7.12 eV/atom for $Ti_2N$, $Ti_3N_2$, $Ti_4N_3$ and TiN, respectively).

2. High temperature endurance. Bulk titanium carbide and nitride exhibit melting points over 3000 K [18,19]. Thermal fluctuations generate thermal strains in the structures, which occur at elastic energy costs. To train MLIPs to simulate TiC and TiN structures at 3000 K, we use the general elastic energy formula [20]:

$$U = U_0 + \frac{1}{2}C_{11}\epsilon_{xx}^2 + \frac{1}{2}C_{22}\epsilon_{yy}^2 + C_{12}\epsilon_{xx}\epsilon_{yy} + 2C_{66}\epsilon_{xy}^2 , \tag{1}$$

where $U_0$ denotes the energy of the optimized structure, $C_{ij}$ are the elastic constants of the material, $\epsilon_{ii}$ represents the strain along the $i$-axis, and $\epsilon_{xy}$ denotes the shear strain. Equation (1) is simplified to define the elastic energy cost due to thermal strains by:

$$U = U_0 + \frac{1}{2}\alpha_T \epsilon_{xx}^2, \tag{2}$$

where $\alpha_T$ is a parameter whose value is chosen to enable the structures to withstand thermal strains at 3000 K. Following a series of tests, we obtained $\alpha_T = 1142$ N/m.

This step uses ten ReaxFF-optimized strained TiC and TiN structures with incremental steps of 0.2% strain, ranging from -1% to +1%. Each structure's energy, from equation (2), is provided, but not its stress tensor or forces on atoms. The 0% structure, used in the initial training step, is not included.

3. Elastic properties. The final step in the MLIP training process involves reproducing the elastic constants for all structures. To accomplish this, the tensile strained structures from the previous step are employed. Their energies are given by equation (2) with $\alpha_T = C_{11}$, and with values of $C_{11}$ obtained from or close to DFT calculations (see Ref. [9]). The values of $C_{11}$ used here are 144.2, 240.3, 336.4 and 400.5 N/m for $Ti_2C$, $Ti_3C_2$, $Ti_4C_3$ and TiC, respectively, and 158.6, 264.0, 370.0 and 416.5 N/m for $Ti_2N$, $Ti_3N_2$, $Ti_4N_3$ and TiN, respectively.

At the completion of each step, the resulting MLIPs are tested to ensure they meet the step's objective. Similar results to DFT calculations mean the MLIPs can be used as initial force fields for the next step. After the third step, only the structural and elastic properties tests need to meet these criteria. The maximum temperature the structures can withstand is documented. The supplementary information contain the final trained MLIPs and basic scripts to run simple MD simulations in LAMMPS.

Four distinct protocols of MD simulations were employed. Energy minimization and in-plane elastic constants were obtained following the methodology described in Ref. [20]. Thermal expansion was obtained through 200 ps simulations at specific temperatures, where the system and box sizes equilibrate under zero pressure. The temperature range is from 200 to 500 K, in steps of 50 K. The equilibrium sizes at each temperature are taken as averages of the last 10 ps of each simulation. The stress-strain simulations entailed the application of a tensile strain rate of $10^{-4}$ ps$^{-1}$ along one direction of the structure, while maintaining a temperature of 300 K and allowing the transversal direction to relax in conjunction with the atom positions. In all dynamic simulations, the timestep was set to 0.1 fs.

Upon obtaining the $C_{ij}$ for all structures, the following equations can be used to calculate their Young's modulus, $E$, the shear modulus, $G$, the linear compressibility, β, and the Poisson's ratio, ϑ, [20]:

$$E = \frac{C_{11}C_{22} - C_{12}^2}{C_{22}}, G = \frac{1}{4}(C_{11} - 2C_{12} + C_{22}), \vartheta = \frac{C_{12}}{C_{22}} \text{ and } \beta = \frac{C_{22} - C_{12}}{C_{11}C_{22} - C_{12}^2}. \tag{3}$$

**RESULTS AND DISCUSSION**

The first set of significant results pertain to the structural parameters, lattice parameter, $a_0$, and thickness of all titanium MXenes obtained with the MLIPs. The trained MLIPs to simulate titanium carbide (nitride) MXenes and the bulk are herein denoted MTP1 (MTP2). Table **1** shows the results for $a_0$ and thickness from the simulations with the trained MLIPs. Despite the use of structures optimized by ReaxFF potential, they are in excellent agreement with the values obtained from DFT calculations. Due to the restriction on the number of references, those concerning the values from DFT calculations can be found in Table 1 of Ref. [9].

**Table 1.** Lattice parameter, $a_0$ and the thickness, $t$, both given in Å, for all MXenes obtained from MD simulations using the MTP1 and MTP2 MLIPs, and the values from DFT calculations from the literature (references for DFT values are given in Ref. [9]).

| Structure | MLIP | $a_0^{DFT}$ | $a_0^{MD}$ | $t^{DFT}$ | $t^{MD}$ |
|---|---|---|---|---|---|
| $Ti_2C$ | MTP1 | [3.00 – 3.04] | 3.06 | [2.23 – 2.31] | 2.56 |
| $Ti_3C_2$ | MTP1 | [2.91 – 3.10] | 3.06 | [4.60 – 4.69] | 5.07 |
| $Ti_4C_3$ | MTP1 | [3.07 – 3.09] | 3.06 | [7.14 – 7.16] | 7.57 |
| $Ti_2N$ | MTP2 | [2.98 – 3.03] | 2.96 | [2.27 – 2.31] | 2.27 |
| $Ti_3N_2$ | MTP2 | [3.00 – 3.07] | 2.96 | [4.60 – 5.12] | 4.62 |
| $Ti_4N_3$ | MTP2 | 2.99 | 2.96 | 7.36 | 7.02 |

Table **2** shows the results of $C_{ij}$ from the simulations using trained MLIPs, which again show excellent agreement with DFT values. $C_{22}^{MD}$ was calculated to test the expected similarity to $C_{11}^{MD}$, since the in-plane symmetry of each MXene was not assumed. There is a minor discrepancy between DFT and MD results for $C_{12}$ of $Ti_4C_3$ and $C_{66}$ for $Ti_3C_2$, but the rest of the results are in excellent agreement.

**Table 2.** Elastic constants, $C_{ij}$, in N/m, for all MXenes obtained from MD simulations using the MTP1 and MTP2 MLIPs, and the values from DFT calculations from the literature (references for DFT values are given in Ref. [9]).

| Structure | MLIP | $C_{11}^{DFT}$ | $C_{11}^{MD}$ | $C_{12}^{DFT}$ | $C_{12}^{MD}$ | $C_{22}^{MD}$ | $C_{66}^{DFT}$ | $C_{66}^{MD}$ |
|---|---|---|---|---|---|---|---|---|
| $Ti_2C$ | MTP1 | [130 – 151] | 139 | [32 – 37] | 36 | 138 | 58 | 51 |
| $Ti_3C_2$ | MTP1 | [219 – 253] | 235 | 40 | 56 | 233 | 107 | 89 |
| $Ti_4C_3$ | MTP1 | [312 – 366] | 329 | 49 | 77 | 327 | 130 | 126 |
| $Ti_2N$ | MTP2 | [150 – 154] | 153 | 41 | 40 | 153 | – | 56 |
| $Ti_3N_2$ | MTP2 | 263 | 260 | – | 65 | 260 | – | 97 |
| $Ti_4N_3$ | MTP2 | 369 | 373 | – | 90 | 374 | – | 142 |

The results are finalized with the Young's modulus, $E$, the shear modulus, $G$, the linear compressibility, β, and the Poisson's ratio, ϑ, of all MXenes in Table **3**. The findings are in excellent agreement with those derived from DFT calculations.

**Table 3.** Elastic quantities, $E$, β, ϑ and $G$, and the linear thermal expansion coefficient (TEC) of all MXenes obtained from MD simulations using the MTP1 and MTP2 MLIPs, and the values from DFT calculations from the literature (references for DFT values are given in Ref. [9]). $E$ and $G$ are given in N/m, β is given in units of $10^{-3}$ (N/m)$^{-1}$ and TEC in units of $10^{-6}$ K$^{-1}$.

| Structure | MLIP | $E$ | β | ϑ | $G$ | TEC |
|---|---|---|---|---|---|---|
| $Ti_2C$ | MTP1 | 129 | 5.7 | 0.263 | 51 | 8.49 |
| | DFT | [119 – 143] | [5.4 – 6.3] | [0.230 – 0.270] | [49 – 54] | – |
| $Ti_3C_2$ | MTP1 | 221 | 3.4 | 0.242 | 89 | 7.80 |
| | DFT | [207 – 248] | [3.4 – 3.7] | [0.154 – 0.241] | [83 – 105] | – |
| $Ti_4C_3$ | MTP1 | 311 | 2.5 | 0.235 | 126 | 7.57 |
| | DFT | [305 – 355] | [2.4 – 2.8] | [0.159 – 0.250] | [123 – 132] | – |
| $Ti_2N$ | MTP2 | 141 | 5.2 | 0.263 | 56 | 10.7 |
| | DFT | [123 – 143] | [5.1 – 5.3] | [0.270 – 0.271] | [55 – 56] | – |
| $Ti_3N_2$ | MTP2 | 243 | 3.1 | 0.249 | 97 | 7.98 |
| | DFT | 238 | 2.3 | 0.265 | 97 | – |
| $Ti_4N_3$ | MTP2 | 351 | 2.2 | 0.241 | 142 | 7.09 |
| | DFT | 347 | 2.2 | 0.250 | 138 | – |

Regarding thermal stability, the trained MLIPs could simulate MXenes up to 700 K and bulk structures up to 1000 K, a limitation that will be overcome in future MLIP versions.

Despite these limitations, two additional dynamic tests yielded satisfactory results. First, the estimation of the linear thermal expansion coefficient (TEC) is addressed. The literature presents the values of the linear TEC of TiC and TiN as $6.4 \times 10^{-6}$ K$^{-1}$ [21] and $9.9 \times 10^{-6}$ K$^{-1}$ [22], respectively. The results from the MD simulations with MTP1 and MTP2 are 6.8 and $5.3 \times 10^{-6}$ K$^{-1}$, thereby indicating that, at least, the TEC of TiC is in good agreement with experimental values. The values of the TEC of MXenes are shown in Table 3, and represent the first estimates of these quantities, as the extant literature on this property is lacking.

The second test involves studying the stress-strain of all structures simulated with trained MLIPs (see Figure 1). There are several studies on the stress-strain of some titanium MXenes. One major observation is that the thickness of MXene directly affects its capacity to withstand strains [23,24]. Our results agree with these studies. The stress values observed in stress-strain simulations of titanium carbide MXenes by Borysiuk, Mochalin, and Gogotsi [25] and Billah et al. [23] are of the same order as ours. The sole DFT study of the stress-strain of the $Ti_3C_2$ MXene [26] reports an exceptionally high rupture strain compared to our findings and those of other studies [23-25].

The Young's modulus values of 503 and 617 GPa for TiC and TiN structures, respectively, can be extracted directly from the stress-strain curves shown in Figure **1**. The experimental values are 449 [27] and 445 [28] GPa, which indicates that the trained MLIPs results provide good agreement with experimental results at 300 K.

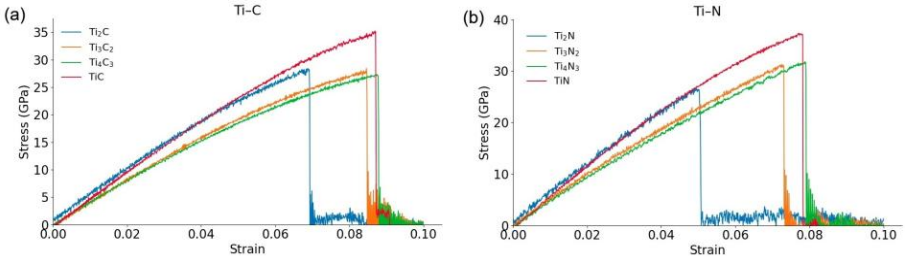

**Figure 1.** Stress-strain curves of titanium carbide (a) and nitride (b) MXenes and bulk.

## CONCLUSION

The preliminary results of a major project of developing effective MLIPs to simulate titanium carbide and titanium nitride MXenes were presented. The MLIP training process, which employed structures not obtained from DFT calculations, was shown to produce remarkable outcomes. Since no MD potential or force field can simulate, at the same time, all titanium MXenes and their corresponding bulk carbide or nitride, our simple yet effective MLIPs are shown to be the first to do so. They produce excellent structural results with values of lattice parameter, thickness, and elastic constants comparable to those of DFT calculations.

These are not the only positive outcomes of the present MLIPs. As recently reported in the literature [29], the development of MLIPs and other AI-based force fields for classical MD simulations should be precise not only to simulate static properties, but also to reproduce dynamical properties. Bearing this in mind, a training strategy has been devised to obtain an MLIP capable of simulating, at least, the bulk structures at temperatures close to their melting points. Our dynamic tests demonstrated that this strategy was ineffective in reproducing the thermal stability of titanium carbide and nitride at temperatures exceeding 1000 K. However, the trained MLIPs were effective in reproducing the experimental value of the TEC of titanium carbide at 300 K, and half the experimental value of the TEC of titanium nitride. The order of magnitude is, at the very least, reasonable. The stress-strain results obtained using our trained MLIPs also demonstrated good agreement with similar studies in the literature, including the verification of the agreement between the extracted Young's modulus at 300 K and that of experiments.

In light of the favorable outcomes derived from our MLIP training methods, we considered that our MLIPs can be used as first approximations to perform tests and initial simulations of the structure and dynamical behavior of large MXenes. Nevertheless, meticulous training strategies will be employed to improve the present MLIPs so as to reproduce more deep physical properties like the phonon spectra of the system [30]. This will be subject of future publications.

## ACKNOWLEDGMENTS

This work used resources of the John David Rogers Computing Center (CCJDR) in the Gleb Wataghin Institute of Physics, University of Campinas.

## FUNDING

This work was supported by the Brazilian Agency CNPq-Brazil (Grant number 302009/2025-6); São Paulo Research Foundation (FAPESP) (Grant number #2024/14403-4); and Fundação de Apoio ao Ensino, Pesquisa e Extensão – FAEPEX/UNICAMP (Grant number #3423/25).

## AUTHOR CONTRIBUTION

L.F.V.T. conceived the idea and performed the training and computational simulations and calculations. A.F.F. and L.F.V.T. analyzed the results. A.F.F wrote the original draft and L.F.V.T. reviewed and edited the draft. A.F.F supervised the work and acquired funding. A.F.F. and L.F.V.T. read and approved the final manuscript.

## CONFLICT OF INTEREST STATEMENT

On behalf of all authors, the corresponding author states that there is no conflict of interest.

## DATA AVAILABILITY STATEMENT

Data available on reasonable request from the authors.